\documentclass{iopart} 

\newcommand{\NBI}{$^1$ QUANTOP, Niels Bohr
  Institute, University of Copenhagen, Blegdamsvej 17, 2100
  Copenhagen, Denmark}
\newcommand{\IQIS}{$^2$ Institute for Quantum Information Science, University of Calgary, 2500 University Dr. NW, Calgary, Alberta, T2N 1N4, Canada}

\usepackage[utf8]{inputenc} 
\usepackage[english]{babel} 
\usepackage{csquotes} 
\usepackage{microtype} 
\usepackage{units} 

\usepackage{hyperref} 
\hypersetup{colorlinks=true, %
  linkcolor=blue, %
  citecolor=blue, %
  filecolor=blue, %
  urlcolor=blue, %
  pdftitle=Tomography of a single polariton state, %
  pdfauthor=NBI, %
  pdfsubject=, %
  pdfkeywords=} %

\usepackage{amssymb} 
\usepackage{amstext} 
\usepackage{amsbsy} 
\usepackage{setstack}

\usepackage{graphicx} 
\graphicspath{{images/}{./}}

\newcommand{\ket}[1]{\left|#1\right\rangle}
\newcommand{\bra}[1]{\left\langle #1\right|} 
\newcommand{\braket}[2]{\langle #1\vert #2 \rangle} 

\newcommand{\up}{\uparrow}
\newcommand{\down}{\downarrow}
\newcommand{\var}[1]{\mathrm{var} \left( #1 \right)} %

\makeatletter
\newcommand{\svnnote}[1]{\@ifundefined{svnid}{}{\textcolor{magenta}{\tiny{#1}}} }
\makeatother

\usepackage[
backend=bibtex8,  
eprint=true, 
sorting=none, 
bibstyle=numeric,
citestyle=numeric, 
defernums=true, 
maxnames=99,
firstinits=true, 
autocite=plain 
]{biblatex}

\DeclareFieldFormat[article]{volume}{\textbf{#1}} 
\DeclareFieldFormat[article]{number}{\textbf{#1}} 

\renewbibmacro*{addendum+pubstate}{\printfield{addendum}} 
\bibliography{ref2} 

\DeclareNameAlias{editor}{default}
\DeclareNameAlias{translator}{default}

\AtEveryBibitem{
  \iffieldundef{doi}{}{\clearfield{eprint}} 
  \iffieldundef{isbn}{}{\clearfield{url}} 
} 

\renewbibmacro*{in:}{} 

\DeclareFieldFormat{titlecase}{\MakeTitleCase{#1}}
	
\newrobustcmd{\MakeTitleCase}[1]{%
  \ifthenelse{\ifcurrentfield{booktitle}\OR\ifcurrentfield{booksubtitle}%
    \OR\ifcurrentfield{maintitle}\OR\ifcurrentfield{mainsubtitle}%
    \OR\ifcurrentfield{journaltitle}\OR\ifcurrentfield{journalsubtitle}%
    \OR\ifcurrentfield{issuetitle}\OR\ifcurrentfield{issuesubtitle}%
    \OR\ifentrytype{book}\OR\ifentrytype{mvbook}\OR\ifentrytype{bookinbook}%
    \OR\ifentrytype{booklet}\OR\ifentrytype{suppbook}%
    \OR\ifentrytype{collection}\OR\ifentrytype{mvcollection}%
    \OR\ifentrytype{suppcollection}\OR\ifentrytype{manual}%
    \OR\ifentrytype{periodical}\OR\ifentrytype{suppperiodical}%
    \OR\ifentrytype{proceedings}\OR\ifentrytype{mvproceedings}%
    \OR\ifentrytype{reference}\OR\ifentrytype{mvreference}%
    \OR\ifentrytype{report}\OR\ifentrytype{thesis}}
    {#1}
    {\MakeSentenceCase{#1}}}
\usepackage[normalem]{ulem}	
\usepackage{color}

\begin{document}

\title[Tomography of a single polariton state]{Towards quantum
  state tomography of a single polariton state of an atomic ensemble}

\author{ S.L.~Christensen$^1$, J.B.~Béguin$^1$, H.L.~Sørensen$^1$, E.~Bookjans$^1$, D.~Oblak$^2$, J.H.~Müller$^1$, J.~Appel$^1$ and E.S.~Polzik$^1$}
\address{\NBI} 
\address{\IQIS} 

\ead{polzik@nbi.ku.dk}

\begin{abstract}
  We present a proposal and a feasibility study for the creation and quantum state tomography of a single polariton state of an atomic ensemble.
  The collective non-classical and non-Gaussian state of the ensemble
  is generated by detection of a single forward scattered photon. The
  state is subsequently characterized by atomic state tomography
  performed using strong dispersive light-atoms interaction followed
  by a homodyne measurement on the transmitted light. The proposal is
  backed by preliminary experimental results showing projection noise
  limited sensitivity and a simulation demonstrating the feasibility
  of the proposed method for detection of a non-classical and
  non-Gaussian state of the mesoscopic atomic ensemble. This work
  represents the first attempt of hybrid discrete-continuous variable
  quantum state processing with atomic ensembles.
\end{abstract}

\pacs{03.65.Wj, 32.80.Qk, 42.50.Dv}


\maketitle

\section{Introduction}
Atomic ensembles have emerged as a highly efficient medium for
light-matter quantum interfaces \cite{Hammerer2010}.  Mapping of
squeezed states of light onto an ensemble
\cite{Georgiades1995,Hald1999}, generation and retrieval of single
excitations \cite{Lukin2003,Choi2008,Albert2011}, entanglement of two
ensembles \cite{Julsgaard2001,Krauter2011,Chou2005} and quantum
sensing \cite{Leroux2010,Napolitano2011,Wasilewski2010} have all been
demonstrated experimentally. The development of interfaces between
light and atomic ensembles have so far followed two separate paths
which until now have been pursued independently. Operating with either
discrete excitations and single photon detection or with continuous
Gaussian states and homodyne measurements. Within the latter approach
tomography of spin squeezed atomic states by quantum non-demolition
(QND) interactions with light has been developed
\cite{Fernholz2008,Louchet-Chauvet2010,Takano2009}.

In close analogy to methods used in photonic systems
\cite{Ourjoumtsev2007,Neergaard-Nielsen2006}, the hybrid approach of
combining discrete excitations, such as number states, and
measurements in the continuous variable domain opens up new venues in
quantum state engineering with atomic ensembles.  Among them are
hybrid quantum repeaters and generation of Schrödinger cat states
\cite{Brask2010}, weak quantum measurements \cite{Simon2011} and
heralded quantum amplification for precision measurements
\cite{Brunner2011}.  

In the following, we report on the first steps towards bridging the
gap between the discrete and continuous variable approaches with
atomic ensembles. Starting from a collective photon scattering as in
Duan et al. \cite{Duan2001}, we present a method to produce and
directly characterize a highly non-classical (negative Wigner
function) and non-Gaussian collective (entangled) atomic state.  Using
interference between different atomic spin-wave modes allows us to
demonstrate its properties. In particular, we will demonstrate how a
single atomic excitation changes the projection
noise statistics of a macroscopic atomic cloud containing over
$100\,000$ atoms fundamentally.

A central point in our proposal is that both the creation and
characterization is performed \emph{directly} in the atomic ensemble. This
has several advantages compared to other methods: First, the direct
creation inside an atomic ensemble with a long coherence time results
in an heralded state that is readily available for on-demand use. Secondly,
in comparison to protocols where the quantum state of the atomic
ensemble is created by mapping of a photonic state on to the ensemble
\cite{Choi2008,eisaman05:_elect,matsukevich20042} our method does not
suffer from the inherent loss mechanism in the transfer of a quantum
state between light and atoms. Third in in comparison to converting the single atomic
excitation back into into the optical domain for further characterization \cite{MacRae2012},
preserving the state inside
the memory presents perspectives for further manipulation, e.g.
improved heralding scaling by shelving the excitation into a third
atomic level, Gaussian state manipulation by dispersive
QND-measurements and feedback, as well as linear optics equivalent
operations between different spin-wave modes.

\section{Generation of a single polariton state}
\label{sec:idea}
\begin{figure}
  \centerline{\includegraphics[keepaspectratio,
    width=\textwidth]{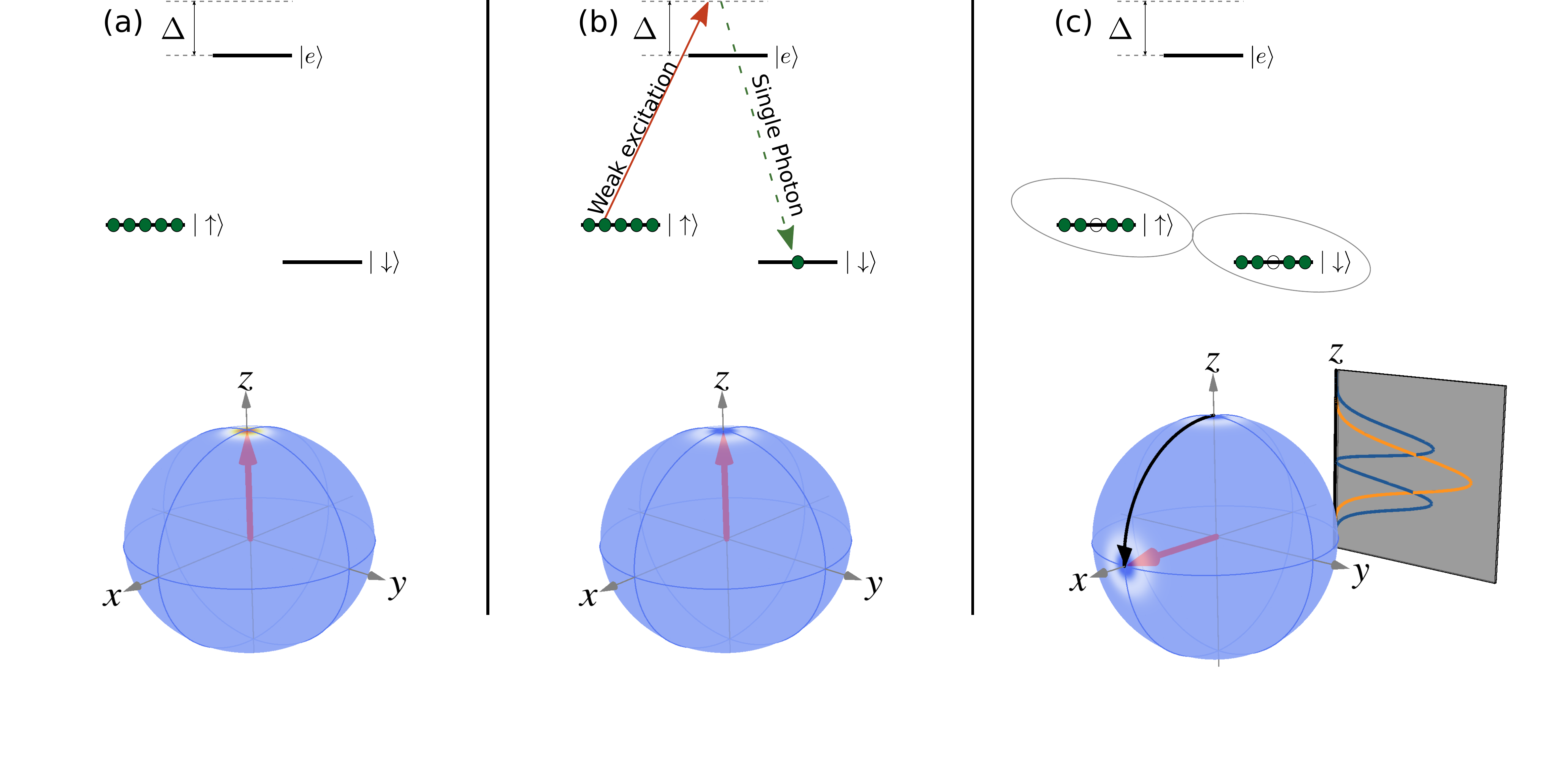}}
  \caption{The atomic levels and the collective Bloch sphere (as
    described by Dowling et al.\cite{Dowling1994}) at different stages
    of the protocol (a) Atoms are prepared in $\ket{\Psi_0}$ and a
    weak blue detuned excitation pulse is applied (b) detection of a
    forward scattered Raman photon signals the generation of a single
    collective excitation in the ensemble (c) a microwave $\pi/2$
    pulse rotates the state $\ket{\Psi_1}$ into the equatorial plane
    of the Bloch sphere, the insert shows a plot of the non-Gaussian
    (Gaussian) marginal distribution (probability density of
    $\hat{J}_z$-measurement outcomes) of the $\ket{\Psi_1'}$
    ($\ket{\Psi_0'}$) state after a rotation to the equatorial plane
    in blue (orange).}
  \label{fig:idea}
\end{figure}

We consider a system of $N_a$ spin-$1/2$ particles described by the
states $\ket{\uparrow}$ and $\ket{\downarrow}$ which in the experiment
are the two hyperfine ground states of a Cesium (Cs) atom, also known
as the clock-levels. Initially all atoms are prepared in the
$\ket{\uparrow}$ state, see \fref{fig:idea}a, such that the quantum state
of the ensemble can be written as the product state
\begin{equation}
  \label{eq:vacuum}
  \ket{\Psi_0} = \bigotimes_{l=1}^{N_a} \ket{\uparrow}_l = \ket{\uparrow\uparrow\ldots\uparrow\uparrow}.
\end{equation}
Next, a weak off-resonant excitation pulse is sent into the ensemble,
and conditioned on the detection of one photon scattered forward via a
Raman process, one spin flip occurs in the collective polariton state
of the ensemble. This detection event heralds the preparation of a
single excitation of a collective atomic zero-transverse-momentum spin
wave - a single polariton, see \fref{fig:idea}b. The success probability
of this forward scattering is kept low, such that the probability to
forward-scatter two or more photons is negligible. The state of the
ensemble \cite{Duan2001} becomes
\begin{equation}
  \label{eq:singleexcitationstate}
  \ket{\Psi_1} = \hat{a}^{\dagger} \ket{\Psi_0}
  = \frac{1}{\sqrt{N_a}} \sum \limits_{l=1}^{N_a} \ket{\uparrow\uparrow\ldots \uparrow \smash{\underbrace{
        \downarrow}_{\makebox[0pt]{\text{$l$-th atom} } } } \uparrow \ldots \uparrow\uparrow},
\end{equation}
\vspace{0.2cm}
where we have defined the symmetric creation operator
\begin{equation}
  \hat{a}^{\dagger} = \frac{1}{\sqrt{N_a}} \sum_{l=1} ^{N_a} \ket{\down}_l\bra{\up}_l.
\end{equation}

To describe the atomic ensemble we define pseudo spin operators for
each individual atom
\begin{eqnarray}
  \hat{j}_x &\equiv \frac{1}{2} \Bigl( \ket{\down} \bra{\up} + \ket{\up} \bra{\down} \Bigr), \\
  \hat{j}_y &\equiv -\frac{i}{2}\Bigl( \ket{\down} \bra{\up} - \ket{\up} \bra{\down} \Bigr), \\
  \hat{j}_z &\equiv \frac{1}{2} \Bigl( \ket{\up} \bra{\up} - \ket{\down} \bra{\down} \Bigr),
\end{eqnarray}
which obey angular-momentum like commutation relations. From these we
define symmetric (under particle exchange) collective ensemble
operators by summing the individual atomic operators as
\begin{equation}
  \hat{J}_i = \sum _{l=1} ^{N_a} \hat{j}_i^{(l)} \quad \text{for} \: i=x,y,z.
\end{equation}
where $N_a$ is the total number of atoms in the ensemble and
$\hat{j}_i^{(l)}$ acts on the $l$-th atom.

A subsequent microwave $\pi/2$-pulse rotates this collective entangled
state, see \fref{fig:idea}c,  such that it becomes an even superposition of
$\ket{\uparrow}$ and $\ket{\downarrow}$ given as

\begin{eqnarray}
  \label{eq:singleexcitationstate2}
  \ket{\Psi_1 '} &=\hat R_{\pi/2} \ket{\Psi_1} \\
  &= \frac{1}{\sqrt{N_a}} \sum \limits_{l=1}^{N_a} \ket{++ \ldots+ \smash{\underbrace{
        -}_{\makebox[0pt]{\text{$l$-th atom} } } } + \ldots +  +},
\end{eqnarray}
\vspace{0.2cm}
where the microwave $\pi/2$ pulse is described by the operator
\begin{equation}
  \hat R_{\pi/2} =  \sum \limits_{l=1}^{N_a} \left ( \ket{+}_l\bra{\uparrow}_l + \ket{-}_l\bra{\downarrow}_l \right ),
\end{equation}
and the coherent superposition states are
\begin{equation}
  \ket{\pm} \equiv \frac{\ket{\uparrow} \pm \ket{\downarrow} }{\sqrt{2}} .
\end{equation}
To show that the state $\ket{\Psi_1'}$ has a non-Gaussian
population-difference between the two hyperfine levels we calculate
the probability to find $n$ atoms in the $\ket{\downarrow}$ state and
the remaining $\left(N_a-n\right)$ atoms in the $\ket{\uparrow}$ state. This is
done by calculating the overlap between the single excitation state,
$\ket{\Psi_1 '}$, and the general $n$-th excited state given by
\begin{equation}
  \ket{\Psi_n}={ N_a \choose n} ^{-1/2} \sum_{\mathrm{permutations}} \ket{\smash{\underbrace{
        \uparrow}_{\text{$N_a - n$}  } } \smash{\underbrace{
        \downarrow}_{\text{$n$} } } }
\end{equation}
we find
\begin{eqnarray}
  \left| \braket{\Psi_n}{\Psi_1'} \right|^2 
  &= 2^{-N_a} { N_a \choose n } \frac{4}{N_a} \left(n-\frac{N_a}{2}\right)^2 \\
  &\stackrel{N_a\to \infty}{\approx} 2 x^2 \sqrt{\frac{2}{\pi N_a}} e^{-x^2}, 
  \label{eq:2}
\end{eqnarray}
whereas the standard coherent spin state (CSS) $\ket{\Psi_0'}$ has
\begin{eqnarray}
  \left| \braket{\Psi_n}{\Psi_0'} \right|^2 
  &= 2^{-N_a} { N_a \choose n }  \\
  &\stackrel{N_a\to \infty}{\approx}
  \sqrt{\frac{2}{\pi N_a}} e^{-x^2},
  \label{eq:3}
\end{eqnarray}
with $x=\sqrt{\frac{2}{N_a}} \left( n-\frac{N_a}{2}\right)$.  From
this it is clear that, \emph{conditioned} on whether a single photon
is detected or not, the probability distributions for outcomes of
$\hat J_y$- and $\hat J_z$-measurements will be profoundly different,
see insert on \fref{fig:idea}c, i.e. the respective probability
for a $\hat J_z$-measurement to give $\left (n-N_a/2\right)$ as outcome is $P(J_z = n-N_a/2)
= \left| \braket{\Psi_n}{\Psi_{0\text{ or }1}'} \right|^2$.

In the ideal case of perfect tomography of a pure single polariton
excitation the population difference follows the marginal distribution
of a Wigner function with a single excitation, a $n=1$ Fock-state,
given by equation \eref{eq:2}. In a realistic case of finite
efficiency of both the state preparation and state detection we show
that the marginal distribution still retains its non-classical and
non-Gaussian features, as discussed in detail in
\sref{sec:implementation}.

Before we continue to the experimental implementation we introduce the
effective quantum efficiency of the quadrature measurement,
$\epsilon$. This is equivalent to the efficiency of the state
tomography which is based on the quadrature measurements. These are
done with a dispersive QND-probing method introduced in
\cite{Saffman2009} and discussed in detail in
\cite{Louchet-Chauvet2010, Appel2009}. We use two
probes, each of which interacts with the $\ket{\uparrow}$
($\ket{\downarrow}$) state exclusively and which therefore does not
cause Raman transitions to the other clock state $\ket{\downarrow}$
($\ket{\uparrow}$), respectively. Each of these probes measures the atomic state
dependent optical phase shift, which we then convert into quadrature
values \cite{Kiesel2012}. By a noise scaling analysis we discriminate
light shot noise and technical noise against the atomic noise. With
this we quantify the efficiency as
\begin{equation}
  \epsilon = \frac{\var{\text{CSS}(N_a)} - \var{\text{CSS}(N_a=0)} }{ \var{\text{CSS}(N_a)}}.
\end{equation}
Here $\text{CSS}(N_a)$ denotes the differential phase shift measured
on a coherent spin state $\ket{\Psi_0'}$ with $N_a$ atoms. Note that
$\var{\text{CSS}(N_a=0)}$ (the optical phase fluctuations measured in
the absence of atoms) contains both shot noise and technical noise.
The contribution of atomic noise to $\var{\text{CSS}(N_a)}$ scales
quadratically with the number of photons in the probe, whereas the
shot noise contribution only scales linearly \cite{Hammerer2010}. For
tomography applications it is not necessary to preserve the
non-destructiveness of the QND probing -- that is to preserve the
coherence between the states $\ket{\uparrow}$ and $\ket{\downarrow}$.
Therefore, neglecting technical noise, a probe of unlimited strength
can in principle be used, and the quantum efficiency of the
tomographic measurement will approach unity.

\section{Implementation}
\label{sec:implementation}
As described in \sref{sec:idea}, the Wigner function of the single
polariton state has both a non-Gaussian marginal distribution and a
negativity and is therefore highly non-classical. Over the past years
several experiments have shown atomic state tomography
\cite{Leroux2010, Fernholz2008, Appel2009, Schmied2011} in which the
characterization of such a state could be performed.  In the following
we will present a detailed discussion on how a single polariton state
can be created and characterized. We will focus on the specific
experimental configuration described in \cite{Louchet-Chauvet2010,
  Appel2009}.

\subsection{Experimental considerations}
\begin{figure}
  \centerline{\includegraphics[keepaspectratio,
    width=\textwidth]{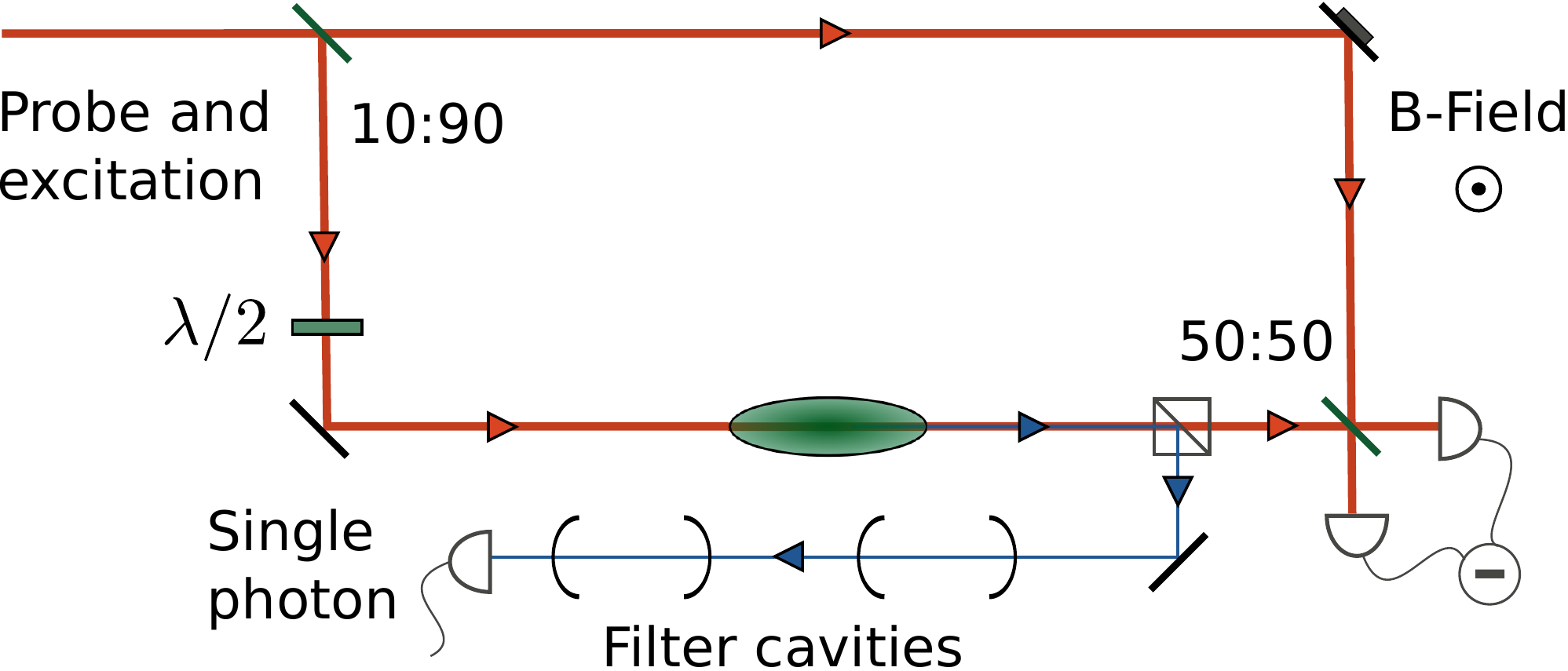}}
  \caption{Sketch of the experimental setup: The atomic ensemble is
    trapped in a far off-resonant dipole trap (not shown). A single
    polariton state is generated by sending a weak excitation pulse
    through the atomic cloud. The detection of a forward scattered
    single photon heralds the creation of the desired state.
    Polarization and frequency filtering is applied to avoid the
    detection of undesired excitation photons using a polarizing beam
    splitter and two Fabry-Perot filter cavities. To characterize the
    atomic state we measure the atomic state dependent optical phase
    shift imprinted on a light pulse by the atoms. }
  \label{fig:setup}
\end{figure}

The experiment is based on an ensemble of approximately $10^5$ Cs
atoms. We consider the two-level system formed by the clock levels,
i.e. $\ket{\downarrow} \equiv \ket{F=3,m_F = 0}$ and $\ket{\uparrow}
\equiv \ket{F=4,m_F = 0}$. As the excited state, used to couple the
two clock states, we use the $\ket{e} \equiv \ket{F'=4, m_F ' =+1}$
state of the $\unit[852]{nm}$ D2 transition.  The atomic ensemble is
situated in a far-off resonant dipole trap which is placed in one arm
of a Mach-Zehnder interferometer (MZI), see \fref{fig:setup}.  To
obtain the probability distribution of $\hat{J_z}$-measurement
outcomes, we detect the \emph{atomic state dependent} optical
phase-shift imprinted on light passing through the atomic ensemble
using a dual-color dispersive QND measurement as described in
\sref{sec:idea}.
To create the single polariton state given by equation
\eref{eq:singleexcitationstate} we propose the following procedure;
initially all atoms are prepared in the $\ket{\uparrow}$ state by a
combination of microwave pulses, optical pumping and purification pulses
\cite{Windpassinger2008a}. The collective excitation is created by a
weak excitation pulse, blue detuned by $\Delta \approx
\frac{3}{2}\Gamma$ from the $\ket{\uparrow} \rightarrow \ket{e}$
transition. The detuning is chosen such that the optical depth is less
than unity, (assumed an on resonant optical depth of $\text{OD}
  \approx 10$) which means that all atoms will have about the same
chance to scatter a photon. If this condition is fulfilled the
excitation will be collective, i.e.  shared almost uniformly by the
whole ensemble.

\begin{figure}
  \centerline{\includegraphics[keepaspectratio,
    width=0.5\textwidth]{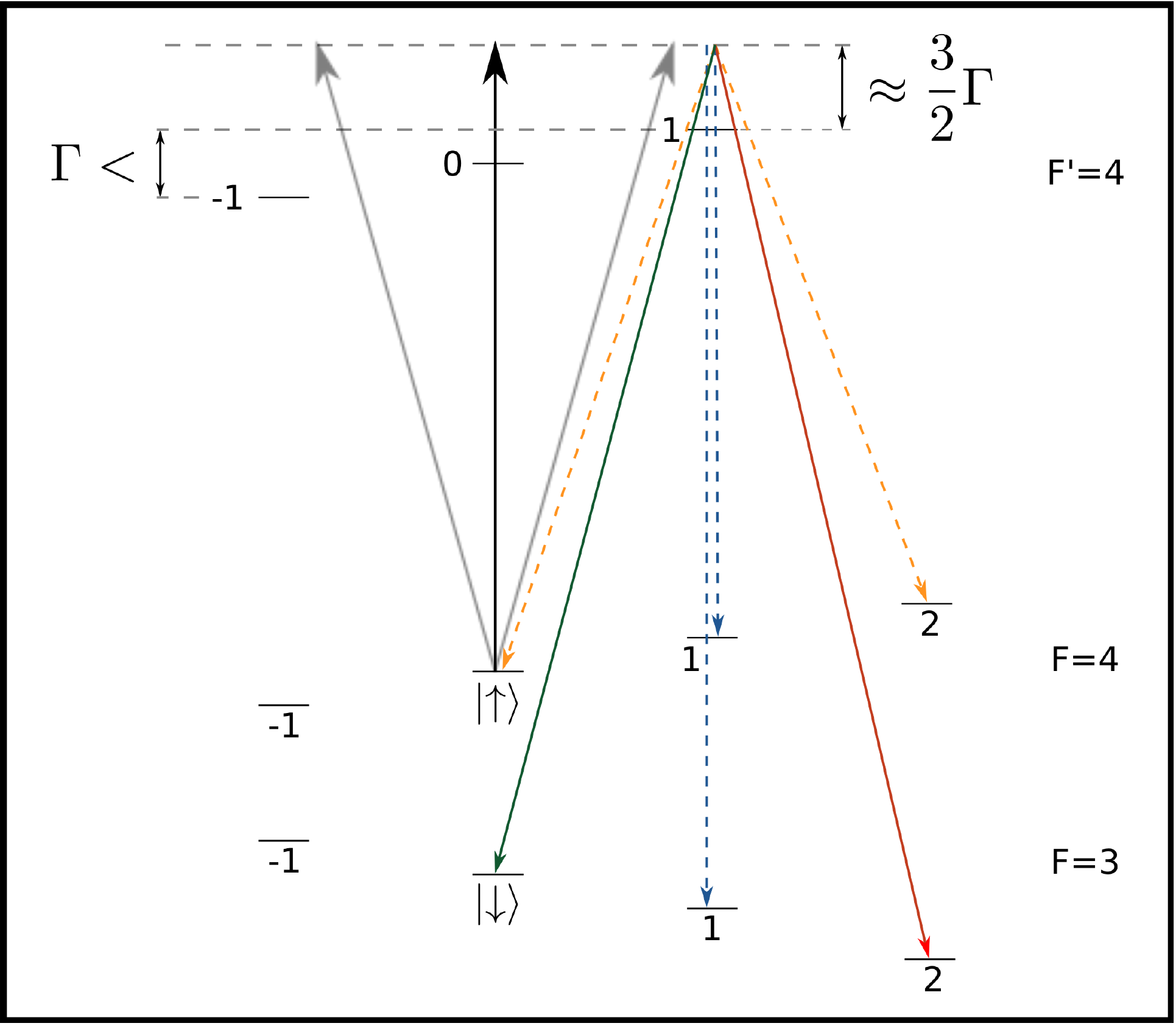}}
  \caption{Cs level structure and excitation scheme: upwards arrows indicate
  the excitation light and the downward arrows
    indicate decay channels. Light corresponding to the dashed arrows is filtered out by either
    polarization (blue) or frequency (orange). Solid arrows denote decays
    that \emph{cannot} be filtered out. Note that the
    decay to $\ket{F=3, m_F=2}$ leads to the unavoidable inefficiency
    of the state generation as discussed in the text. }
  \label{fig:nonGaussianLevels}
\end{figure}

To obtain the best possible spatial overlap between the probe beams
and the excitation beam with the atomic ensemble all beams arrive at
the MZI input via the same optical fiber, whose output is carefully
overlapped with the atomic sample. This nearly perfect overlap of the
light beams obtained by using the same optical fiber comes with the
constraint that the polarization of the excitation beam and the probe
beams are now identical. With the quantization axis (in $z$-direction,
defined by the magnetic bias-field) chosen orthogonal to the
propagation direction of the beams ($y$-direction), we can only
address the atoms with $\pi$- or $x$-polarized light (i.e. linearly
polarized orthogonal to $\pi$-polarization).  In order to obtain the
highest efficiency of the atomic tomography it is preferable to use
$\pi$-polarized light \cite{Louchet-Chauvet2010}.  This combined with the complicated
multi-level structure of the Cs atoms puts several constraints on the
excitation and probe geometry and necessitates the application of a
strong magnetic bias field as discussed in the following.  Because of
the symmetry of the $m_F=0$ atomic wave functions, anti-Stokes photons
on the Raman transition $\ket{\uparrow} \to \ket{F'=4} \to
\ket{\downarrow}$ \emph{cannot} be emitted into the forward direction
in the absence of an external magnetic field. Since for both $x$- and
$\pi$-polarized excitation beams, the Clebsch-Gordan coefficients
corresponding to the relevant dipole matrix elements are such, that
excitation can only occur into superpositions of excited state
Zeeman-sublevels for which the same-polarization decay into the
respective other $\ket{F,m_F=0}$ hyperfine ground state interferes
destructively in the absence of magnetic fields. The applied bias field is approximately
$\unit[20]{Gauss}$ which shifts the $\ket{F'=4, m_F' = -1}$ state out of
 resonance by several line widths, see~\fref{fig:nonGaussianLevels}.
 This choice combined with the fact that
the $\ket{\uparrow} \rightarrow \ket{F' = 4, m_F'=0}$ transition is
forbidden by selection rules allows us to achieve the required
selective excitation to the $\ket{e}$ state. From the $\ket{e}$ state
the atom can undergo spontaneous emission through six possible decay
channels, see~\fref{fig:nonGaussianLevels}.

We now apply different filtering methods such that we only detect
photons corresponding to an atom decaying via the $\ket{e} \rightarrow
\ket{\downarrow}$ transition, since this projects the ensemble into
the the single polariton state given by equation
\eref{eq:singleexcitationstate}.  Photons with $\pi$-polarization are
suppressed by a polarizing beam splitter. The harder task is to filter
out photons that are $x$-polarized and have a frequency corresponding
to the $\ket{e} \to \ket{F=4}$ manifold. Photons with these properties
originate from both the excitation beam and the spontaneous decay. We
thus need to reject photons with a frequency difference of
approximately $\unit[9]{GHz}$. This is done with two cascaded
Fabry-Perot filter cavities, see \fref{fig:setup}.  This way we filter
out all photons except the ones originating from the undesired decay
channel $\ket{e} \rightarrow \ket{F=3, m_F = 2}$. This decay channel
produces forward scattered photons which have the same polarization as
the desired photon, the frequency difference between them is in the
$\unit{MHz}$ range (the Zeeman splitting of the ground state), which
makes it experimentally very hard to selectively reject them. Note
that the branching ratios favor the preferred decay by a factor of
$\approx 4$.

With the single photon detected, we proceed in a fashion similar to
J.~Appel et al. \cite{Appel2009}. We apply a $\pi/2$-pulse and perform
the dispersive QND probing of the atomic population difference in
$\ket{\uparrow}$ and $\ket{\downarrow}$.  Repeating this procedure
several thousand times we characterize the distribution of
$\hat{J_z}$-measurements for the prepared atomic state.  Note that due
to the rotational symmetry of the Wigner function of both the single
polariton- and the coherent spin-state the $\pi/2$ rotation can be
performed without paying attention to the phase of the microwave pulse
(which defines the axis of rotation on the Bloch sphere).

\subsection{Test of tomography noise performance}
\begin{figure}
  \centerline{\includegraphics[keepaspectratio,angle=-90,
    width=0.75\textwidth]{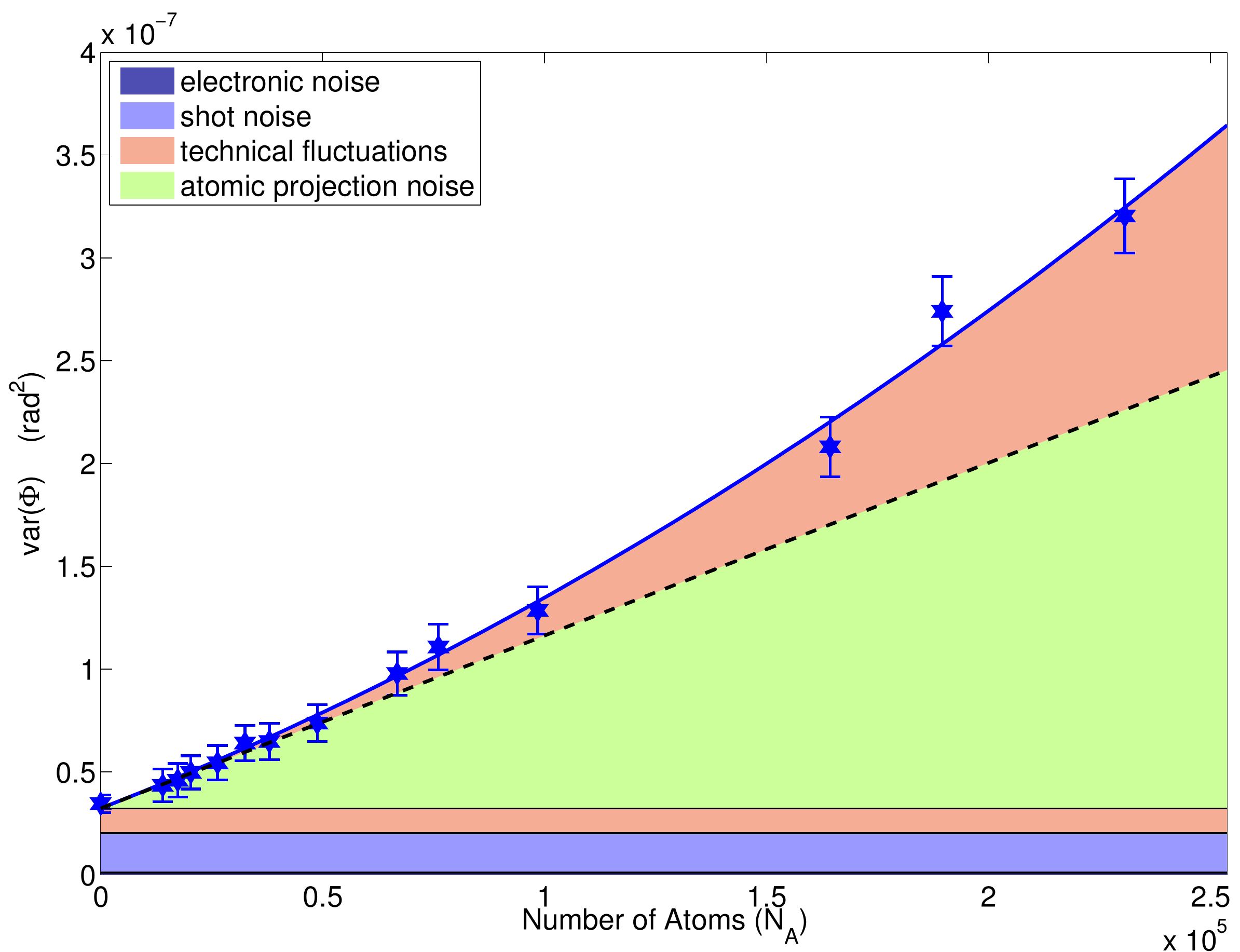}}
  \caption{The measured variance of the probe phase as a function of
    atom number, with different noise contributions discriminated by
    noise scaling analysis. Blue line: general quadratic fit to data
    (blue stars with statistical errors). The atomic projection noise
    (green area) scales linearly with $N_a$.  Technical fluctuations
    (red areas) originate both from noise in the QND measurement as
    well as imperfections in the CSS preparation.}
  \label{fig:NoiseScaling}
\end{figure}
The major experimental challenge is concerning the characterization of
the atomic state via quantum tomography, since this requires the
measurement of the population difference ($\hat{J_z}$) in the
experimental apparatus to be projection noise limited. We have checked
our noise sensitivity in the experimental configuration with a
bias-field of $B \approx \unit[20]{Gauss}$, corresponding to a
splitting between the $\ket{e}$ and $\ket{F'=4, m_F' = -1}$ states of
several line widths.

We start by preparing a CSS of the atomic ensemble (a product state of
each atom being in $\ket{+}$) \cite{Appel2009} and then measure the
population difference between the two clock-levels $\ket{\uparrow}$
and $\ket{\downarrow}$. Repeating this several thousands times and
performing a noise scaling analysis using the same method as in
\cite{Appel2009}, we attribute the measured fluctuations to different
origins, see \fref{fig:NoiseScaling}. We see a predominantly linear
dependence of the atomic noise on the atom number (green area) -- a clear signature
of the required projection noise limited sensitivity. This certifies
sufficient performance of the initial state preparation, the quality
of our microwave source and the long term stability of our setup.

The next part of the experiment is to detect the single photon,
distinguishing and filtering out photons originating from the
excitation beam and undesired decay channels.  Detection of any of
these photons would result in false positive ``clicks'', i.e.  we
detect a photon but the atomic ensemble is not prepared in the desired
single excitation state $\ket{\Psi_1}$. Such ``clicks'' could be due
to dark counts, leakage of the probe-, trap- or excitation-beams.
Since the state generation is based on the detection of a single
photon with a low success probability, false positives essentially mix
in realizations prepared in the vacuum state (zero polaritons).

To describe this, we consider the actual state prepared (after the $\pi/2$ microwave pulse) conditioned on
a ``click'' as a classical mixture of the single excitation state
$\ket{\Psi_1'}$ which is obtained with a probability $p$, and the CSS, $\ket{\Psi_0'}$,
obtained with a probability $(1-p)$:
\begin{equation}
  \label{eq:1}
  \hat \rho = p \ket{\Psi_1'}\bra{\Psi_1'} + (1-p) \ket{\Psi_0'}\bra{\Psi_0'}
\end{equation}
In order to show that the outlined procedure is suitable for our
experimental apparatus we perform an analysis based on a simulation
using the relevant experimental parameters.

\section{Analysis}
\begin{table}
  \caption{ 
    \label{tab:vacuum} Probability for photo-counts of different origins. We have only kept the most
    dominant processes, i.e. probabilities greater then $\unit[0.05]{\%}$.}
  \lineup
  \begin{indented}
  \item[] \begin{tabular}{@{}lll} \br
      Origin of photo-count & Created state & Probability (\%) \\
      \mr
      Dark counts & $\ket{\Psi_0}$ &$ \0 8.7$ \\
      Leakage of excitation pulse &  $\ket{\Psi_0}$ & $\0 0.6$ \\
      Decay via $\ket{e} \rightarrow \ket{F=3,m_F=2}$ & $\ket{\Psi_0}$ &	$19.1$ \\
      Decay via $\ket{e} \rightarrow \ket{\down}$ & $\ket{\Psi_1}$ & $71.5$	 \\
      \br
    \end{tabular}
  \end{indented}
\end{table}
\begin{figure}
  \centerline{\includegraphics[keepaspectratio,
    width=1\textwidth]{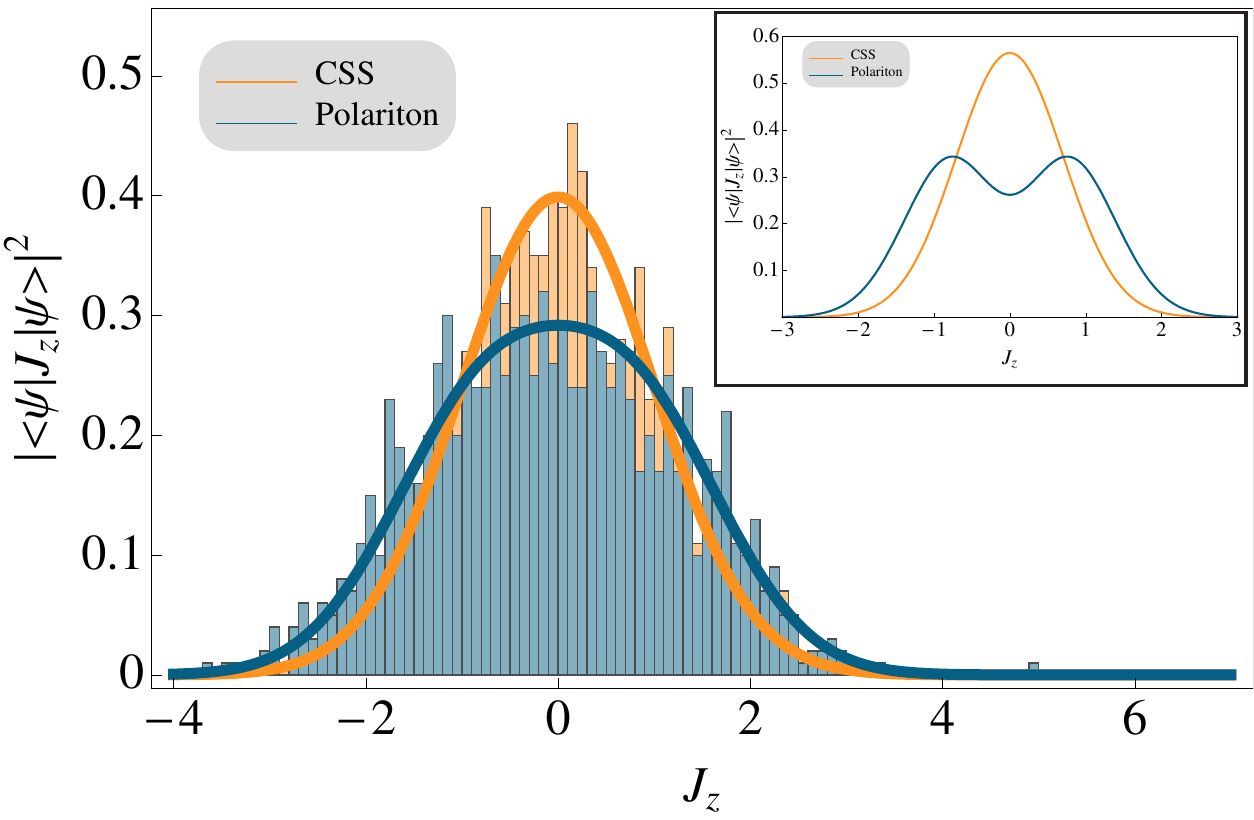}}
  \caption{Calculated distributions for $\hat{J}_z$-measurements outcomes of the
    created single polariton state (blue) and the vacuum state
    (orange), taking into account the finite tomography efficiency.
    The histogram is based on 1000 randomly drawn samples from each
    distribution. Insert: Marginal distributions \emph{not} including the effects of
      a non-unity detection efficiency. Note that the
    less-than-unity tomography efficiency washes away the non-Gaussian
    feature, making it hard to distinguish the two histograms by eye.}
  \label{fig:Marginal}
\end{figure}
To quantify the non-Gaussian character of the prepared state we
calculate the expected probability distributions for our experimental
parameters and perform a simulation similar to Dubost et al.
\cite{Dubost2012}.

In order to estimate the purity of the single polariton excitation
(the probability $p$ introduced in the previous section), we assume
that the excitation pulse is so weak that the probability of producing
a forward-scattered photon is $\unit[5]{\%}$. To reach this scattering
probability we require $1.35 \cdot 10^{4}$ photons in a $\unit[10]{\mu
  s}$ long excitation pulse. The polarization filtering is done with a
polarizing beam splitter cube with a rejection of $1:7\cdot 10^{3}$.
For the frequency filtering we obtain rejections of $1:5\cdot 10^7$ by
using two cascaded filter cavities, each with a transmission of
$\unit[80]{\%}$.  The mode-matching overlap of the single photon
detection mode and the atomic state tomography mode is taken to be
$\unit[75]{\%}$. With this we estimate the probabilities
of the photo-count originating from a specific decay channel, a dark
count of the detector or from the leakage of other light sources (e.g.
the excitation pulse) in the experiment.  The probabilities of the
dominating processes (those exceeding $\unit[0.05]{\%}$) are shown in
\tref{tab:vacuum}.

From the probability given in \tref{tab:vacuum} we calculate the
probability distributions for $\hat{J}_z$-measurements. The created state is
modeled as a statistical mixture of a CSS and a ideal, noise free
single excitation state weighted with their corresponding
probabilities, see equation \eref{eq:1}, a plot is shown in the insert on \fref{fig:Marginal}.
The efficiency of the atomic state tomography is reduced due to the
extra noise added by the readout procedure: We use light as the
meter-system and therefore the shot noise of light will also have a
contribution.  Following the approach outlined in \cite{Appel2007} we
model this independent Gaussian noise as an extra admixture of the vacuum
state to the state to be analyzed. In effect this reduces the
non-Gaussianity of the $\hat{J}_z$ distribution detected by the
dispersive measurements as shown in the main panel of figure
\ref{fig:Marginal}.

As pointed out in \cite{Dubost2012}, the detection of a non-Gaussian
marginal distribution is an experimental challenge. One of the main
reasons for this is that we can only estimate the underlying
probability distributions by a finite number of experimentally
acquired samples. To illustrate this, we draw 1000 samples from the
distributions for the dispersive $\hat{J}_z$-measurements. These samples are binned and plotted as a histogram in
\fref{fig:Marginal}. It is clear from the histograms that
distinguishing the two sets of sampled data and thus detecting the
non-Gaussianity and possibly the non-classicality of the quantum state
is experimentally challenging. Several methods have been developed in
order to quantify the non-classicality \cite{Kot2012, Vogel2000},
we note that these methods define the term ``non-classical'' in different ways. The
criterion by Kot et al. does not implicitly assume quantum mechanics,
whereas the less stringent Vogel criterion does: A state described by
equation~\eref{eq:1} is non-classical for all $p$ according to the
Vogel criterion \cite{Lvovsky2002}, whereas only states with $p > 0.5$
are non-classical by the definition of Kot. et al.

\begin{figure}
   \centerline{\includegraphics[keepaspectratio,
     width=1\textwidth]{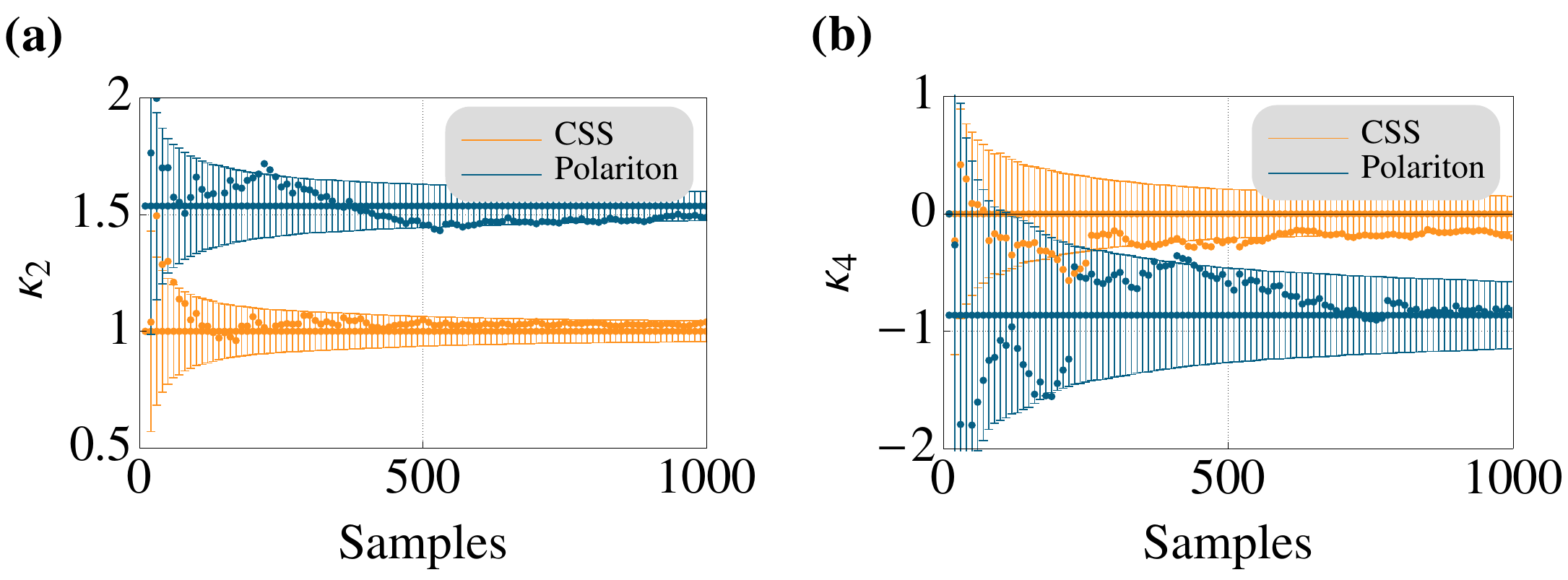}}
   \caption{Plot of the cumulants $\kappa_2$ and $\kappa_4$ as a
     function of the sample size. Dotted points correspond to data from one
     realization simulation.}
   \label{fig:Cumulants}
 \end{figure}

In the following we present a simulation based on the ideas in
\cite{Dubost2012} where the non-Gaussianity is quantified in terms of
the statistical cumulants. From the known probability distributions we
draw random samples and use these to calculate the second and fourth
cumulants, $\kappa_2=\mu_2$ and $\kappa_4=\mu_4 - 3\mu_2^2$, for
varying sample sizes, $\mu_i$ denotes the $i$-th statistical moment.
Repeating this 1000 times allows us to calculate the corresponding
standard deviations, see \fref{fig:Cumulants}. We start by considering
the second order cumulant, which is just the variance of the
distribution: From \fref{fig:Cumulants}a we note that only around 250
samples are needed to distinguish between the distribution of the
single polariton state and the vacuum (CSS). We thus expect that
careful investigation of $\kappa_2$ will allow us to clearly classify
our data as originating from either the expected single polariton
state or the CSS under the assumptions that these are the only
possible distributions. In order to verify that our state is
non-Gaussian we need to observe a non-zero forth cumulant $\kappa_4$.
We expect to require approximately 500 samples to clearly distinguish
the observed probability distribution form the Gaussian one of the CSS
with $\kappa_4 = 0$. The performed simulations are based on realistic
parameters, many already experimentally verified, and we can thus
conclude that our proposed implementation is experimentally feasible.

 \section{Conclusion}
 In this paper we have introduced a hybrid method based on discrete
 excitations and continuous measurements which allows us to generate
 and characterize a single polariton state of an atomic ensemble. This
 state has non-Gaussian marginal distributions of the Wigner function
 and is non-classical. We have presented a detailed proposal for the
 experimental creation and detection of the single polariton state.
 The proposal is backed by a simulation for experimental valid
 parameters together with preliminary results showing the feasibility
 of the proposal. This work is a step towards implementing a hybrid
 approach to quantum information processing with atomic memories as
 well as a contribution to the ongoing research into the foundations
 of quantum mechanics.

 \ack We thank A.~Sørensen and E.~Kot for helpful discussions
 regarding the state preparation and quantification of
 non-classicality and A.~Louchet-Chauvet, J.~J.~Renema, and
 N.~Kjærgaard for contributions in the earlier stages of the
 experiment. The presented work has been supported by the DARPA project QUASAR, EU project
 Q-ESSENCE and ERC grant INTERFACE.

 \printbibliography[heading=bibintoc]

\end{document}